\documentclass[10pt,conference]{IEEEtran}
\IEEEoverridecommandlockouts

\usepackage[ruled,lined,linesnumbered,vlined]{algorithm2e}

\SetCommentSty{mycommfont}
\DontPrintSemicolon 

\SetKwProg{Fn}{Function}{:}{}

\usepackage{amsmath}
\usepackage{amsfonts}
\usepackage{amsthm}
\usepackage{balance}
\usepackage{enumitem}
\usepackage{graphicx}
\usepackage{listings}
\usepackage{multicol}
\usepackage{multirow}
\usepackage{subcaption}
\usepackage{url}
\usepackage{xspace}
\usepackage{xcolor}
\usepackage{tcolorbox}
\usepackage{colortbl}

\usepackage{tikz}
\usepackage{pgfplots}

\usetikzlibrary{shapes}
\usetikzlibrary{shapes.geometric}
\usetikzlibrary{arrows.meta, positioning}

\newcommand{\eg}{\mbox{\textit{e.g.}}\xspace}
\newcommand{\etal}{\mbox{\textit{et al.}}\xspace}
\newcommand{\etc}{\mbox{\textit{etc.}}\xspace}
\newcommand{\ie}{\mbox{\textit{i.e.}}\xspace}
\newcommand{\vs}{\mbox{\textit{vs.}}\xspace}

\definecolor{BlindColorTolOne}{HTML}{332288}
\definecolor{BlindColorTolTwo}{HTML}{117733} 
\definecolor{BlindColorTolThree}{HTML}{44AA99}
\definecolor{BlindColorTolFour}{HTML}{88CCEE}
\definecolor{BlindColorTolFive}{HTML}{DDCC77}
\definecolor{BlindColorTolSix}{HTML}{CC6677} 
\definecolor{BlindColorTolSeven}{HTML}{AA4499}
\definecolor{BlindColorTolEight}{HTML}{882255}

\definecolor{BlindColorWongOne}{HTML}{000000} 
\definecolor{BlindColorWongTwo}{HTML}{E69F00}
\definecolor{BlindColorWongThree}{HTML}{56B4E9}
\definecolor{BlindColorWongFour}{HTML}{009E73}
\definecolor{BlindColorWongFive}{HTML}{F0E442}
\definecolor{BlindColorWongSix}{HTML}{0072B2} 
\definecolor{BlindColorWongSeven}{HTML}{D55E00}
\definecolor{BlindColorWongEight}{HTML}{CC79A7}

\definecolor{mygreen}{HTML}{02818a}

\mathchardef\mhyphen="2D

\newcounter{FindingCounter}

\newcommand{\myparagraph}[1]{
  \vspace*{0.04cm}
  \noindent \textit{\textbf{#1.}}\quad
}

\SetKwData{AlgTrue}{true}
\SetKwData{AlgFalse}{false}
\SetKw{AlgBreak}{break}
\SetKw{AlgContinue}{continue}

\newcounter{myUniqueIdCounter}

\makeatletter
\newcommand{\myGetOrAssignID}[1]{%
  \ifcsname myMap@#1\endcsname%
    \csname myMap@#1\endcsname%
  \else%
    \stepcounter{myUniqueIdCounter}%
    \expandafter\xdef\csname myMap@#1\endcsname{\themyUniqueIdCounter}%
    \themyUniqueIdCounter%
  \fi%
}
\makeatother

\newcommand{\anonymizedId}[1]{\ifx\useAnonymizedId\undefined%
  #1%
\else%
  \myGetOrAssignID{#1}%
\fi}

\newcommand{\proj}{\text{PeepholeBench}\xspace}

\newcommand{\gemini}{\text{Gemini CLI}\xspace}
\newcommand{\codex}{\text{Codex CLI}\xspace}
\newcommand{\claude}{\text{Claude Code}\xspace}
\newcommand{\agy}{\text{Antigravity CLI}\xspace}

\newcommand{\geminiThreeFlash}{\text{Gemini 3 Flash}\xspace}
\newcommand{\geminiThreeFiveFlash}{\text{Gemini 3.5 Flash}\xspace}
\newcommand{\gptFivePointFour}{\text{GPT 5.4}\xspace}
\newcommand{\gptFivePointFourMini}{\text{GPT 5.4 mini}\xspace}

\newcommand{\sonnetFourPointSix}{\text{Sonnet 4.6}\xspace}

\newcommand{\llm}{\text{LLM}\xspace}
\newcommand{\llms}{\text{LLMs}\xspace}
\newcommand{\llvm}{\text{LLVM}\xspace}
\newcommand{\instcombine}{\text{InstCombine}\xspace}
\newcommand{\ir}{\text{IR}\xspace}
\newcommand{\irs}{\text{IRs}\xspace}
\newcommand{\opt}{\textsc{opt}\xspace}
\newcommand{\llvmmca}{\text{llvm-mca}\xspace}
\newcommand{\alivetwo}{\text{Alive2}\xspace}
\newcommand{\xEightSix}{\text{x86\_64}\xspace}
\newcommand{\aarchSixFour}{\text{AArch64}\xspace}
\newcommand{\llvmlit}{\text{llvm-lit}\xspace}
\newcommand{\filecheck}{\text{FileCheck}\xspace}

\newcommand{\prCollectionStartDate}{Sept.~1, 2025\xspace}

\newcommand{\numCollectedIssues}{\text{27}\xspace}
\newcommand{\numCollectedPRs}{\text{25}\xspace}
\newcommand{\numIssues}{\text{21}\xspace}
\newcommand{\numPRs}{\text{19}\xspace}

\newcommand{\percentMaxInconclusiveRate}{10\%\xspace}
\newcommand{\numMutIterations}{5\xspace}
\newcommand{\numMutantsPerIteration}{200\xspace}
\newcommand{\numMutAttemptFactor}{100\xspace}
\newcommand{\numMutMax}{1,000\xspace}

\newcommand{\numLabelCases}{\text{160}\xspace}
\newcommand{\numThemes}{\text{8}\xspace}
\newcommand{\kappaScore}{\text{0.834}\xspace}

\newcommand{\evalRqIValidityPassAtThreeOfGptFivePointFour}[1]{79#1\xspace}
\newcommand{\evalRqIValidityPassAtThreeOfSonnetFourPointSixAndGptFivePointFourMini}[1]{68#1\xspace}
\newcommand{\evalRqIValidityPassAtThreeOfGeminiThreeFlash}[1]{63#1\xspace}

\newcommand{\evalRqIProfitabilityPassAtThreeOfGeminiThreeFlash}[1]{42#1\xspace}
\newcommand{\evalRqIProfitabilityPassAtOneOfGeminiThreeFlash}[1]{28#1\xspace}

\newcommand{\evalRqIIAgentValidityLowest}[1]{97.0#1\xspace}
\newcommand{\evalRqIIAgentValidityHighest}[1]{98.5#1\xspace}
\newcommand{\evalRqIIHumanValidity}[1]{98.4#1\xspace}
\newcommand{\evalRqIIGeminiValidity}[1]{97.7#1\xspace}
\newcommand{\evalRqIIGeminiProfitability}[1]{16.5#1\xspace}
\newcommand{\evalRqIIHumanProfitability}[1]{16.4#1\xspace}
\newcommand{\evalRqIIOtherAgentsProfitabilityLow}[1]{8.8#1\xspace}
\newcommand{\evalRqIIOtherAgentsProfitabilityHigh}[1]{9.0#1\xspace}
\newcommand{\evalRqIIGeminiGeneralization}[1]{3.7#1\xspace}
\newcommand{\evalRqIIHumanProfitabilityIssueInvalidCount}{287\xspace}
\newcommand{\evalRqIIHumanInvalidTotal}{288\xspace}
\newcommand{\evalRqIIHumanProfitabilityIssueInvalidPct}[1]{99.7#1\xspace}
\newcommand{\evalRqIIHumanLargeRegressionPct}[1]{15#1\xspace}

\newcommand{\cost}{\text{cost}\xspace}
\newcommand{\cpucycles}{\text{cycles}\xspace}
\newcommand{\microops}{\text{uops}\xspace}
\newcommand{\profit}{\mathcal{P}\xspace}
\newcommand{\shr}{\mathbin{\gg}}

\newcommand{\pr}{\text{PR}\xspace}
\newcommand{\prs}{\text{PRs}\xspace}

\newcommand{\metricB}{\Scale[0.8]{\textsf{B}}\xspace}
\newcommand{\metricS}{\Scale[0.8]{\textsf{S}}\xspace}
\newcommand{\metricV}{\Scale[0.8]{\textsf{V}}\xspace}
\newcommand{\metricP}{\Scale[0.8]{\textsf{P}}\xspace}

\newcommand{\code}[1]{\Scale[0.9]{\texttt{#1}}}
\pgfplotsset{compat=1.18}

\usepackage{tabularx}
\usepackage{fvextra}
\usepackage[hidelinks]{hyperref}
\usepackage[capitalise,noabbrev]{cleveref}
\usepackage{cite}
\usepackage{amsmath,amssymb,amsfonts}
\usepackage{graphicx}
\usepackage{textcomp}
\usepackage{xcolor}
\usepackage{balance}
\usepackage{booktabs}
\usepackage{multirow}
\usepackage{array}
\usepackage{stfloats}
\usepackage[table]{xcolor}

\definecolor{g1}{RGB}{220,240,220}
\definecolor{g2}{RGB}{170,220,170}
\definecolor{g3}{RGB}{100,200,100}

\definecolor{TSKeywordColor}{HTML}{005EA8}
\definecolor{TSTypeColor}{HTML}{7A3E9D}
\definecolor{TSInstructionColor}{HTML}{C44E00}
\definecolor{TSFlagColor}{HTML}{8A5A00}
\definecolor{TSFunctionColor}{HTML}{008060}
\definecolor{TSNameColor}{HTML}{8B2A5B}
\definecolor{TSConstantColor}{HTML}{4D4D4D}
\definecolor{TSCommentColor}{HTML}{6B7280}

\newcommand{\TSCodeFont}{\fontsize{6.8}{7.4}\selectfont}
\newcommand{\TSCodeInput}[1]{%
  \VerbatimInput[
    commandchars=\\\{\},
    fontsize=\TSCodeFont,
    baselinestretch=1,
    breaklines=true,
    breakanywhere=true,
    rulecolor=\color{black!30}
  ]{#1}%
}

\newcommand{\llvmIssueURL}[1]{\href{https://github.com/llvm/llvm-project/issues/#1}{\##1}}
\newcommand{\llvmPRURL}[1]{\href{https://github.com/llvm/llvm-project/pull/#1}{\##1}}

\newcommand{\anonymizedLlvmPRURL}[2]{\ifx\useAnonymizedId\undefined%
  \llvmPRURL{#1}%
\else%
  \href{https://anonymous.4open.science/pr/#2}{\#anon.#2}%
\fi}

\theoremstyle{definition}
\newtheorem{definition}{Definition}[section]

\newcounter{MyFindingCounter}
\newcommand{\myfinding}[2][]{%
	\begin{tcolorbox}[boxrule=0.1mm,left=2pt,right=2pt,top=2pt,bottom=2pt,before skip=4pt,after skip=4pt]
		\small\textbf{Finding~\refstepcounter{MyFindingCounter}\theMyFindingCounter%
		\ifx\relax#1\relax\else\label{#1}\fi}: #2
	\end{tcolorbox}%
}
\crefname{MyFindingCounter}{Finding}{Findings}
\crefname{paragraph}{\S}{\S}

\newcommand*{\Scale}[2][4]{\scalebox{#1}{$#2$}}%

\makeatletter
\newcommand{\linebreakand}{%
  \end{@IEEEauthorhalign}
  \hfill\mbox{}\par
  \mbox{}\hfill\begin{@IEEEauthorhalign}
}
\makeatother

\usepackage{cleveref} 

\Crefname{algocf}{Algorithm}{Algorithms}
\crefname{algocf}{Algorithm}{Algorithms}

\Crefname{algorithm}{Algorithm}{Algorithms}
\crefname{algorithm}{Algorithm}{Algorithms}

\crefname{appendix}{Appendix}{Appendices}
\Crefname{appendix}{Appendix}{Appendices}

\Crefname{figure}{Figure}{Figures}
\crefname{figure}{Figure}{Figures}

\crefname{listing}{Listing}{Listings}
\Crefname{listing}{Listing}{Listings}

\Crefname{table}{Table}{Tables}
\crefname{table}{Table}{Tables}

\crefname{thm}{Theorem}{Theorems}
\Crefname{thm}{Theorem}{Theorems}

\crefname{equation}{Equation}{Equations}
\Crefname{equation}{Equation}{Equations}

\crefformat{chapter}{\S~#2#1#3}
\crefmultiformat{chapter}{\S\S~#2#1#3}{ and~#2#1#3}{, #2#1#3}{, and~#2#1#3}

\crefformat{section}{\S~#2#1#3}
\crefmultiformat{section}{\S\S~#2#1#3}{ and~#2#1#3}{, #2#1#3}{, and~#2#1#3}

\crefname{definition}{Definition}{Definitions}

\def\BibTeX{{\rm B\kern-.05em{\sc i\kern-.025em b}\kern-.08em
    T\kern-.1667em\lower.7ex\hbox{E}\kern-.125emX}}

\begin{document}

\title{Can Coding Agents Implement Missed Compiler Optimizations? Evaluating LLM Agents on \llvm Peephole Optimizations}

\ifx\useAnonymizedId\undefined
\author{\IEEEauthorblockN{Hongxu Xu}
\IEEEauthorblockA{\textit{Cheriton School of Computer Science} \\
\textit{University of Waterloo}\\
Waterloo, Canada \\
hongxu.xu@uwaterloo.ca}
\and
\IEEEauthorblockN{Chunhao Liao}
\IEEEauthorblockA{\textit{Cheriton School of Computer Science} \\
\textit{University of Waterloo}\\
Waterloo, Canada \\
chunhao.liao@uwaterloo.ca}
\linebreakand
\IEEEauthorblockN{Xintong Zhou}
\IEEEauthorblockA{\textit{Cheriton School of Computer Science} \\
\textit{University of Waterloo}\\
Waterloo, Canada \\
x27zhou@uwaterloo.ca}
\and
\IEEEauthorblockN{Chengnian Sun}
\IEEEauthorblockA{\textit{Cheriton School of Computer Science} \\
\textit{University of Waterloo}\\
Waterloo, Canada \\
chengnian.sun@uwaterloo.ca}
}
\else
\author{\IEEEauthorblockN{Anonymous Author(s)}}
\fi

\maketitle
\thispagestyle{plain}
\pagestyle{plain}

\begin{abstract}
Coding agents built on large language models
are now capable of patching sizable real-world codebases,
yet whether they can develop compiler optimizations remains an open question.
To study this question, we introduce \proj, an evaluation framework
whose tasks are constructed from real-world
missed peephole optimizations reported against \llvm's
\instcombine pass.
Since missed peephole optimizations are typically fixed with small, localized patches,
they offer a well-scoped but demanding testbed for coding agents:
a correct fix demands rigorous reasoning about program semantics
along with familiarity with optimizer-specific conventions.
\proj derives its tasks from \numIssues resolved \llvm issues and \numPRs merged
pull requests (\prs), supplies agents with only the issue context that existed
before each fix, and assesses the resulting patches for both correctness and profitability.

With \proj, we benchmark state-of-the-art coding agents on fixing missed
peephole optimizations in \llvm's \instcombine pass,
measuring their patches against the corresponding human-written fixes.
We observe a tension between correctness and profitability,
and no agent matches human developers on both dimensions at once.
The dominant failure modes are overly narrow transformations and misuse of
\llvm-specific mechanisms, errors that existing test suites rarely expose.
Together, these results establish \proj as a realistic and
challenging benchmark for coding agents, and suggest future
directions for building agents that can more dependably assist compiler
optimization development.

\end{abstract}

\begin{IEEEkeywords}
coding agents, compiler optimization, \llvm
\end{IEEEkeywords}

\section{Introduction}
\label{sec:intro}
Agents powered by large language models (\llms)
have shown promise on software development
tasks~\cite{jimenez2023swe,yang2024swe,zhang2024autocoderover,wang2024openhands}.
Coding agents, such as
\gemini~\cite{geminicliGeminiDocumentation} and
\claude~\cite{claudeClaudeCode},
can inspect repositories, edit source files, run tests, and produce
non-trivial patches for large software systems~\cite{merrill2026terminalbench,zhou2026featurebench}.
Their ability to implement compiler optimizations, however, remains unclear.
Compiler optimization is an important and challenging software development task~\cite{sasnauskas2017souper,lopes2021alive2},
in which program semantics must be preserved while improving the performance or size of generated code,
and the implementation must follow optimizer-specific conventions~\cite{lopes2015provably,llvmInstCombineContributor,kwon2025optimization}.
In agentic coding benchmarks, a patch is usually considered successful if it passes the test suites~\cite{yang2024swe,merrill2026terminalbench}.
However, for compiler optimizations, passing existing test suites is often insufficient.
A patch may pass the tests by overfitting to the reported
missed optimization cases (under-generalization),
or by generalizing beyond the test cases in ways that are semantically invalid or unprofitable.
Conversely, a patch may implement a correct and profitable optimization yet fail
tests due to producing a structurally different but semantically equivalent result,
such as reordering the operands of a commutative arithmetic operation.

We study coding agents' ability to implement compiler optimizations
through \llvm's \instcombine pass, the middle-end peephole optimization pass
that rewrites \llvm intermediate representation (\ir) within small windows to
simplify programs, reduce generated code, and expose later
optimizations~\cite{xu2026lpo,lopes2015provably}.
Implementing missed peephole optimizations in \instcombine offers
a focused yet realistic testbed for coding agents for three reasons.
First, missed peephole optimizations are frequently reported,
providing a recurring source of real-world optimization tasks.
Second, their fixes are often localized,
making the resulting behavior amenable to focused \ir tests,
mutation-based testing, and semantic validation~\cite{xu2026lpo,lopes2015provably,lopes2021alive2,kwon2025optimization}.
Third, \instcombine serves as a common entry point for new \llvm
contributors~\cite{archiveContributeLLVM} and provides stable coding
guidelines~\cite{llvmInstCombineContributor}, making it a well-specified and
accessible starting point for evaluating coding agents on compiler optimization tasks.
Together, these properties make missed \instcombine optimizations
a natural setting for automation and benchmarking.
Yet the task remains challenging:
plausible patches still require precise reasoning about \ir
semantics, profitability, and interactions with other optimizations~\cite{lopes2015provably,lopes2021alive2}.

\myparagraph{\proj}
We present \proj, an evaluation framework that tasks coding agents with
implementing missed \llvm peephole optimizations.
\proj is constructed from \numIssues closed \llvm issues reporting
missed \instcombine optimizations, alongside the \numPRs merged pull requests (PRs) that
resolve them.
For each task, \proj reconstructs the repository at the \pr's base
commit (the commit immediately preceding the human patch) and provides the agent with
the linked issue descriptions,
available \alivetwo examples (before-and-after \ir pairs illustrating the optimization)~\cite{lopes2021alive2},
and the parts of the \instcombine development guide~\cite{llvmInstCombineContributor} that
are relevant to implementing peephole optimizations.
This setup evaluates agents on the exact same compiler development task
originally faced by human contributors, without exposing the merged patch.

In this paper, we study how far current coding agents are from implementing
compiler optimizations by evaluating three state-of-the-art coding agents with \proj:
\gemini~\cite{geminicliGeminiDocumentation} (\geminiThreeFlash),
\codex~\cite{openaiCodexOpenAI} (\gptFivePointFour and \gptFivePointFourMini),
and \claude~\cite{claudeClaudeCode} (\sonnetFourPointSix).
We first measure how often each agent can produce a
buildable patch that passes the conventional \llvm test suites for missed
\instcombine optimizations. We then go beyond merely passing the test suites: using
mutation-based testing, \alivetwo validation~\cite{lopes2021alive2},
and \llvmmca-based profitability estimation~\cite{llvmLlvmmcaLLVM},
we evaluate whether generated patches achieve
\emph{behavioral validity}, requiring the resulting
rewrite to be both \emph{correct} and \emph{non-regressive},
as formally defined in \cref{def:behavioral-validity}.
We further compare agent-generated patches with accepted human patches
and analyze common failure modes to identify where current agents still
fall short in compiler development.

Our evaluation reveals that no agent matches human experts in
terms of both validity and profitability simultaneously.
\sonnetFourPointSix matches human-expert validity rates
but produces profitable optimizations for
substantially fewer programs.
\geminiThreeFlash inverts this trade-off,
achieving profitability on a range of programs comparable to human-written patches,
but at a modest cost in validity.
We attribute these shortcomings to two recurring limitations.
First, agents tend to overfit to the concrete optimization instance
described in the issue report and fail to generalize beyond it (\cref{par:rq3:under-generalization}).
Second, they mishandle \llvm-specific correctness constraints---such as guarding
single-use values (\cref{par:rq3:one-use-guard}), preserving
undefined-behavior semantics (\cref{par:rq3:flag-misuse}),
and avoiding conflicts with existing optimization rules (\cref{par:rq3:fold-interaction})---that
issue reports leave underspecified and existing test suites often fail to cover.
Together, these findings suggest that agentic development of compiler optimizations
requires better evaluation methodologies, tighter integration of semantic
validation tools into the agent loop, targeted strategies to close generalization and
domain-knowledge gaps, and a rethinking of how agents and human contributors
can complement each other; we discuss these directions in \cref{sec:implications}.

\myparagraph{Contributions}
We make the following contributions:
\begin{itemize}[leftmargin=*]
    \item
    We present and publicly release \proj, a benchmark of \numPRs missed
    \instcombine optimizations derived from real-world \llvm issues and \prs.

    \item We design an evaluation framework on top of \proj that augments
    the existing test suites with mutation-based generation to cover more
    edge cases and assesses behavioral validity to capture both
    the correctness and profitability requirements of compiler optimizations.

    \item We empirically evaluate representative coding agents on \proj,
    comparing their generated patches with human-authored ones.
    We also perform a thematic analysis of failure modes and discuss
    key implications for future coding agents in compiler development.

\end{itemize}

\section{Background}
\label{sec:background}
\begin{figure*}[t]
  \centering
  \begin{minipage}[t]{0.39\textwidth}
  \begin{subfigure}[t]{\textwidth}
    \TSCodeInput{figures/background-instcombine-reported-ir.ts.tex}
    \caption{An \ir and its optimized form, where
      \code{lshr} is a logical right shift,
      \code{trunc} truncates a value to a smaller type,
      and \code{icmp eq} compares integers for equality.}
    \label{fig:background-instcombine-example:reported-ir}
  \end{subfigure}

  \vspace{1ex}

  \begin{subfigure}[t]{\textwidth}
    \TSCodeInput{figures/background-instcombine-test-ir.ts.tex}
    \caption{Another pair of \ir from the test suites, where
      \code{icmp ult} compares integers for unsigned less-than.}
    \label{fig:background-instcombine-example:test-ir}
  \end{subfigure}
  \end{minipage}
  \hfill
  \begin{minipage}[t]{0.59\textwidth}
  \begin{subfigure}[t]{\textwidth}
    \TSCodeInput{figures/background-instcombine-human.ts.tex}
    \caption{Human patch covers both cases.
      The matcher \code{m\_Shr} matches a logical or arithmetic right shift,
      \code{m\_Power2} matches a power-of-two constant,
      \code{m\_LowBitMask} matches a constant of the form $2^n - 1$,
      \code{m\_Value} matches any value,
      \code{countr\_zero} count trailing zeros,
      and \code{replaceInstUsesWith} replaces uses of an instruction (\code{trunc}) with another (\code{icmp eq/ult}).}
    \label{fig:background-instcombine-example:human}
  \end{subfigure}

  \vspace{0.75ex}

  \begin{subfigure}[t]{\textwidth}
    \TSCodeInput{figures/background-instcombine-agent.ts.tex}
    \caption{Agent patch only matches \emph{logical} shifts using \code{m\_LShr} and misses the mask pattern shown in \cref{fig:background-instcombine-example:test-ir}.}
    \label{fig:background-instcombine-example:agent}
  \end{subfigure}
  \end{minipage}
  \caption{
    Missed \instcombine optimization addressed by \llvm \pr~\llvmPRURL{157030}.
    The human patch handles both power-of-two and low-bit-mask patterns;
    the patch generated by \codex handles limited patterns and instructions.
  }
  \label{fig:background-instcombine-example}
\end{figure*}

This section introduces \llvm's \instcombine pass,
detailing how missed peephole optimizations are typically
reported, implemented, and validated.

\subsection{Peephole Optimization in \llvm}

\llvm~\cite{lattner2004llvm} translates programs into a typed intermediate
representation (\ir) and optimizes them through a series of transformation passes.
A common technique within these passes is
\emph{peephole optimization}, which replaces localized instruction patterns with
simpler or more efficient, yet semantically equivalent, code~\cite{mckeeman1965peephole,fischer1991crafting}.
Specifically, \llvm's \instcombine pass serves as a middle-end peephole optimizer,
aggressively applying target-independent simplifications and canonicalizations
directly to the \ir~\cite{lopes2015provably,xu2026lpo,mukherjee2024hydra}.

A \emph{missed peephole optimization} occurs when the compiler fails to perform an
expected, beneficial rewrite.\footnote{
In the remainder of this paper, we use ``rewrite'' to refer to a specific \ir
transformation (especially in negative cases), and ``optimization'' to refer to
the general rule that applies across instances.}
Within the \llvm community, developers typically report these missed optimizations
by providing a minimal \ir example alongside the expected optimized form~\cite{xu2026lpo,mukherjee2024hydra}.
For instance, \Cref{fig:background-instcombine-example:reported-ir} illustrates a real issue
report (\llvmIssueURL{156898}) where the expression $\mathrm{trunc}_{\mathrm{i1}}(4 \shr x)$---which
truncates the shift result to a 1-bit boolean---can be simplified to $x \mathbin{==} 2$.
Since $4$ is $\mathtt{0b100}$ in binary, right-shifting it by $x$ yields a $1$
in the least significant bit if and only if $x \mathbin{==} 2$.
Consequently, the entire truncation expression can be replaced by the direct comparison $x\mathbin{==}2$.

\subsection{Implementing \instcombine Optimizations} \label{subsec:instcombine-development}

Implementing a missed peephole optimization in \llvm's \instcombine pass
typically involves three steps~\cite{mukherjee2024hydra,llvmInstCombineContributor}.
First, a contributor identifies a missed optimization and reports it with a minimal
example (a pair of \ir instruction sequences showing the code before and after the optimization), as seen
in \Cref{fig:background-instcombine-example:reported-ir}.
Second, developers generalize this example into a broader rewrite pattern.
For instance, the human patch in \Cref{fig:background-instcombine-example:human}
generalizes the initial report by using \code{m\_Shr} to capture both logical (\code{lshr})
and arithmetic (\code{ashr}) shifts, while also handling both power-of-two
constants and low-bit masks (values of the form $2^n - 1$).
Third, developers implement the rewrite in \instcombine and validate it with regression tests
via \llvm's integrated tester (\llvmlit)~\cite{llvmLLVMIntegrated}.
This involves writing test files with \filecheck directives~\cite{llvmFileCheckFlexible}
that specify the execution command (\code{RUN}) and the expected optimized \ir
(\code{CHECK}), as shown later in \cref{fig:rq1:cs:161020:input}.
\Cref{fig:background-instcombine-example:test-ir} illustrates such a test case,
verifying that the generalized pattern successfully rewrites $\mathrm{trunc}_{\mathrm{i1}}(15 \shr x)$
to $x < 4$.

\myparagraph{Coding Conventions} To maintain codebase consistency, \llvm strictly enforces its coding conventions.
Proposed patches that fail to adhere to these standards are rejected during code review.
Consequently, contributors are required
to express these rewrites using \llvm's specialized pattern-matching utilities and
helper functions~\cite{llvmInstCombineContributor}. As \Cref{fig:background-instcombine-example:human}
demonstrates, this process relies heavily on abstractions like \code{m\_Shr} (right shift),
\code{m\_Power2} (power-of-two constant), and \code{m\_LowBitMask} (constants of
the form $2^n - 1$).

\subsection{Validating Correctness and Profitability of \ir Rewrites}

An optimization must be both correct and profitable before it can be accepted into \llvm~\cite{llvmInstCombineContributor}.

To guarantee \emph{correctness},
an optimization must preserve the program's observable behavior
whenever the original program is free of undefined behavior.
Manual testing cannot reliably catch the subtle semantic errors that can occur,
such as mishandling
\code{undef} or \code{poison} values~\cite{llvmLLVMLanguage,llvmLLVMUndefined,lee2017taming}
or relying on incorrect preconditions.
Consequently, developers rely on \alivetwo,
an SMT-based verifier that
formally proves that the target \ir refines\footnote{
Refinement is a formal relation that ensures the target's behavior is consistent with
or more defined than that of the source.}
the source \ir under \llvm's semantics~\cite{lopes2015provably,lopes2021alive2}.

To establish \emph{profitability},
an optimization must yield a measurable performance gain.
Even a semantically correct optimization may be rejected
if it increases compiler complexity
without making the generated code meaningfully faster~\cite{llvmInstCombineContributor}.
This profitability can be estimated statically using tools like the \llvm Machine Code Analyzer (\llvmmca),
which predicts the CPU cycles and micro-operations
that a code sequence requires on a given target architecture (e.g., \xEightSix or \aarchSixFour).
These predictions provide concrete evidence to help developers decide
whether an optimization justifies its inclusion~\cite{llvmLlvmmcaLLVM}.

\section{Methodology}
\label{sec:method}
\begin{figure}[t]
  \centering
  \includegraphics[width=\columnwidth]{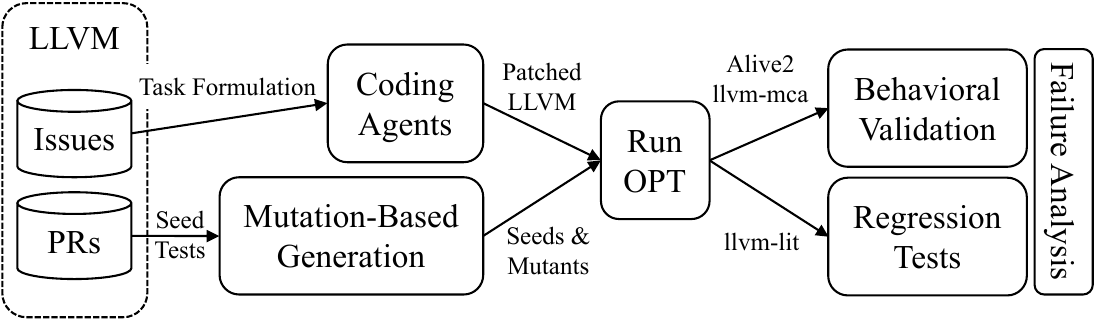}
  \caption{Overview of the \proj{} workflow.}
  \label{fig:workflow}
\end{figure}

We propose \proj, a rigorous evaluation framework designed to assess the
capability of coding agents in implementing missed peephole optimizations within
\llvm's \instcombine pass.
This section details \proj.
\cref{fig:workflow} provides an overview of the end-to-end workflow.

\subsection{Task Formulation}

A task in \proj presents the agent with an \llvm development environment
for one missed \instcombine optimization.
To instantiate the task, we check out the \llvm repository at the \pr's base commit,
\ie, the parent of the merge commit before the human
patch is applied.
We then generate a task-specific prompt file that serves as the
primary specification for the agent.
Upon completion, we collect the agent's implementation as a patch file
using \code{git diff} for downstream evaluation.

The prompt is structured into four functional components:
\begin{itemize}[leftmargin=*]
    \item \textbf{Issue Specification}:
    The titles and descriptions of linked GitHub issues,
    defining the reported missed optimization.

    \item \textbf{Reference \irs}:
    When available, a source--target \ir pair extracted from
    \alivetwo links provided in the issue, serving as a concrete optimization target.

    \item \textbf{Optimization Guidelines}:
    Excerpts from the \instcombine contributor guide~\cite{llvmInstCombineContributor},
    covering core conventions such as profitability, canonicalization, pattern matching, \etc

    \item \textbf{Instructions}:
    Step-by-step task instructions: generalize the optimization, add regression
    tests and confirm the baseline does not optimize them, implement the optimization
    in the relevant files, and verify that the test cases are optimized.
\end{itemize}

\subsection{Data Collection} \label{subsec:data-collection}

We construct \proj from GitHub issues and PRs in the \llvm
repository, mined via the GitHub API. Our pipeline first identifies
closed issues labeled with both \code{instcombine} and
\code{missed-optimization}. For each issue, we retrieve the corresponding
merged PRs created after \prCollectionStartDate.
This cutoff focuses the benchmark on recent development instances,
minimizing potential data leakage during agent training.

For each candidate pull request, we extract metadata including
timestamps, merge and base commits, and linked issues.

\subsection{Mutation-Based Test Generation} \label{subsec:fuzzing}

\begin{table}[tb]
  \caption{Mutation strategies used in test generation.}
  \label{tab:mutators}
  \footnotesize
  \setlength{\tabcolsep}{3pt}
  \begin{tabular}{lp{3.4cm}>{\scriptsize}p{3.1cm}}
    \toprule
    \textbf{Strategy} & \textbf{Description} & \textbf{Example} \\
    \midrule
    boundary-val          & Replace constant with boundary value & \code{add \%x,8} $\to$ \code{add \%x,-1} \\
    fake-use              & Insert \code{llvm.fake.use} & adds \code{fake.use(\%v)} \\
    gep-shape             & Vary constant GEP index to boundary value & \code{gep \%p,4} $\to$ \code{gep \%p,0} \\
    instr-flags           & Toggle instruction flags & \code{add nsw} $\to$ \code{add} \\
    intrinsic-bool        & Flip boolean poison flag of \code{ctlz}, \code{cttz}, and \code{abs} & \code{ctlz(x,1)} $\to$ \code{ctlz(x,0)} \\
    operand-subst         & Replace operand with dominating same-typed value & \code{add a,b} $\to$ \code{add a,a} \\
    operator-subst        & Swap opcode/predicate & \code{add} $\to$ \code{sub} \\
    power-of-two          & Replace $2^k$ with nearby value & \code{and \%x,8} $\to$ \code{and \%x,7} \\
    select-shape          & Swap or merge true arms and false arms & \code{sel c,a,b} $\to$ \code{sel c,b,a} \\
    shift-shape           & Move shift amount to $0$, $1$, or $\mathit{bw}-1$ & \code{shl \%x,3} $\to$ \code{shl \%x,31} \\
    type-subst            & Replace a type with another of same family & \code{i32} $\to$ \code{i64} \\
    vector-shape          & Vary shuffle mask to identity, reverse, or splat & \code{sflvec a, b, <0,1,2>} $\to$ \code{sflvec a, b, <2,1,0>} \\
    \bottomrule
  \end{tabular}
  \par\smallskip
  {\scriptsize\textbf{Note:} \code{nsw}/\code{nuw} = no signed/unsigned wrap; \code{bw} = bit width; \code{sel} = \code{select}; \code{gep} = \code{getelementptr}; \code{sflvec} = \code{shufflevector}.}
\end{table}

We use mutation-based methodology to generate a test suite that exceeds the
coverage of the manually written tests in the original PRs.
Starting from the test cases (\textit{seeds}) modified or added in the \pr,
we apply a suite of targeted mutators (see \cref{tab:mutators}) to produce
pattern-adjacent \ir variants.
The objective is not to maximize random diversity,
but to expose latent gaps in correctness, profitability, and generalization.
Several mutators deliberately target pervasive \instcombine precondition classes,
such as use counts (\code{fake-use}) and instruction flags (\code{instr-flags}).
For instance, the seed test in \cref{fig:background-instcombine-example:test-ir}
only covers logical right shifts (\code{lshr}),
whereas the human-written patch also handles arithmetic right shifts (\code{ashr}).
Applying the \code{operator-subst} mutator,
which replaces an operator with another from the same family,
can automatically cover these critical edge cases.

To generate a diverse set of mutants while bounding runtime,
we maintain a seed pool $\mathcal{S}$ initialized with the original \irs and execute $K$ iterations.
In each iteration, we sample an existing seed and a mutator to produce a new mutant,
which is retained if it is valid
and unique.
Retained mutants are added back to the pool $\mathcal{S}$ at the end of each iteration,
allowing mutations to compose over time.
To bound runtime, an iteration terminates once it reaches a target of $N$ retained mutants
or exceeds an attempt threshold $T$.
The total number of generated mutants per task is at most $K \cdot N$.
We discuss the specific values chosen for these hyperparameters in \cref{sec:experimental-setup:dataset}.

\subsection{Behavioral Validation} \label{subsec:behavioral-validation}

We evaluate agent-generated patches by semantic correctness and architectural
profitability, rather than by syntactic test passing alone.
Standard \llvm regression tests typically compare the \ir optimized
by a patch against a reference \ir (the output of the human-written patch).
Such textual comparisons are prone to false negatives: an agent may produce a
rewrite that differs syntactically from the reference but is semantically
equivalent and equally profitable.
For example, \code{add i8 \%a, \%b} and \code{add i8 \%b, \%a} are
semantically identical, but would fail a strict string comparison.

To address this limitation, we lift validation from syntax to semantics,
using \alivetwo for correctness and \llvmmca for profitability analysis.
Because optimizations often depend on target-specific data layouts
(such as memory alignments),
\llvm regression tests typically execute the optimizer \opt
under multiple configurations.
To rigorously evaluate a patch across all intended architectures,
we extract these configuration arguments from the \code{RUN} directives
embedded in the original test files.
For each configuration, we run the patched \opt to generate the optimized \ir
and then evaluate its behavioral validity.
To provide a rigorous basis for evaluation,
we define cost, profitability, and
behavioral validity as follows.

\begin{definition}[Cost and Profitability]\label{def:profitability}
Let $\cost(I) = (\cpucycles(I), \microops(I))$ denote the estimated cost of an
\ir program $I$, where $\cpucycles$ and $\microops$ are the CPU cycles and
micro-operations reported by \llvmmca, respectively.
For two \ir programs $S$ and $T$, we write $T \preceq S$ if
$$\Scale[0.9]{
    \cpucycles(T) \le \cpucycles(S)
    \;\land\;
    \microops(T) \le \microops(S),
}$$
and $T \prec S$ if
$$\Scale[0.9]{
    T \preceq S
    \;\land\;
    \bigl(\cpucycles(T) < \cpucycles(S)
    \lor
    \microops(T) < \microops(S)\bigr).
}$$
The profitability of $T$ relative to $S$ is classified as:
$$\Scale[0.9]{
\profit(S,T) =
\begin{cases}
    \textsc{Profitable}, & \text{if } T \prec S, \\
    \textsc{Neutral}, & \text{if } T \preceq S \land T \not\prec S, \\
    \textsc{Mixed}, & \text{if } T \not\preceq S \land S \not\preceq T, \\
    \textsc{Regressive}, & \text{if } S \prec T.
\end{cases}
}$$
We say $T$ is \emph{non-regressive} relative to $S$ if $\profit(S,T) \neq \textsc{Regressive}$,
\ie, we do not classify $T$ as regressive if it improves one cost metric even if it worsens another.
\end{definition}

\begin{definition}[Behavioral Validity]\label{def:behavioral-validity}
A rewrite is \emph{behaviorally valid} if and only if it satisfies both:
\begin{itemize}[leftmargin=*]
    \item \textbf{Correctness}:
    The output \ir is semantically equivalent to, or a refinement of, the input
    \ir, as verified by \alivetwo~\cite{lopes2021alive2}.
    \item \textbf{Profitability}:
    The output \ir is \emph{non-regressive} relative to the unoptimized input
    \ir,
    as estimated by \llvmmca using \xEightSix backends in our evaluation.
\end{itemize}
If \alivetwo fails to return a conclusive result (\eg, due to a timeout or
unsupported features) and \llvmmca does not indicate a regression, the
validation result is \emph{inconclusive}.
Inconclusive cases are excluded from all analyses.
\end{definition}

\subsection{Thematic Analysis of Representative Failure Modes} \label{subsec:failure-mode-analysis}

To characterize the limitations of current coding agents,
we conduct an LLM-assisted thematic analysis of representative failure modes~\cite{braun2006using,dai2023llm,qiao2025thematic}.
We collect all test cases in which at least one agent patch is \emph{behaviorally invalid}
or \emph{regressive} relative to the human patch.
Failure modes are grouped into three high-level categories---\emph{profitability issue},
\emph{optimization failure}, and \emph{incorrect transformation}---each
subdivided by \opt error messages,
\alivetwo counterexamples, and \llvmmca cost deltas
(\cref{tab:rq3-agent-failures}).
We then perform stratified sampling, ensuring at least one case per stratum.

For each sampled case, we assemble a diagnostic bundle containing the input \ir,
agent and human patches, \opt outputs, \alivetwo results, \llvmmca cost deltas,
and the rewrite stack trace.
Bundles are grouped by \pr and provided to two independent \llm labelers,
\claude (\sonnetFourPointSix) and \codex (\gptFivePointFour),
each assigning open-coding labels with a written rationale.
Two authors collaboratively map codes
from both labelers to candidate themes,
refining them by splitting or merging,
with each labeler's codes mapped independently,
preventing intentional alignment that would inflate agreement
at the cost of theme validity.
A third \llm agent, \agy (\geminiThreeFiveFlash),
assigns final themes in closed-coding style.
Fleiss's $\kappa$~\cite{fleiss1971measuring}
is computed across three labelers to ensure reliability (see \cref{subsec:rq3} for details),
and the two authors together make the final determination for divergent cases.

\section{Evaluation}
\label{sec:evaluation}
We evaluate the performance of three state-of-the-art coding agents in
implementing missed peephole optimizations
for \llvm's \instcombine pass using \proj.
Our evaluation is guided by three research questions (RQs):
\begin{itemize}[leftmargin=*]
    \item \textbf{RQ1 (Success Rate):}
    To what extent can coding agents produce buildable patches
    that pass existing tests, satisfy behavioral validity,
    and achieve the same level of profitability as human-written patches?
    \item \textbf{RQ2 (Behavioral Comparison):}
    Are patches written by humans always behaviorally valid? Can coding agents find more profitable optimizations than human experts?
    \item \textbf{RQ3 (Failure Modes):} What are the representative failure modes of agent-generated patches, and what do they reveal about the limitations of agents in compiler optimization?
\end{itemize}

\subsection{Experimental Setup}

\subsubsection{Dataset} \label{sec:experimental-setup:dataset}
To mitigate data leakage, we collect only \prs merged after \prCollectionStartDate.
We initially identified \numCollectedIssues \llvm issues for
\instcombine missed optimizations and \numCollectedPRs resolving \prs.
Because \alivetwo can return inconclusive results,
we exclude \prs with $>$\percentMaxInconclusiveRate inconclusive test cases.
We also manually exclude \prs not implementing an \instcombine peephole optimization
(\eg, implementing in the other pass).
This leaves \numIssues issues and \numPRs \prs.
For mutation-based test generation, we use $K=\numMutIterations$ iterations,
targeting $N=\numMutantsPerIteration$ mutants per iteration with
an attempt threshold $T=\numMutAttemptFactor \cdot N$,
yielding up to $\numMutMax$ mutants per \pr.
\cref{tab:dataset} summarizes the dataset,
detailing modified lines of code (excluding tests),
seed test cases extracted, and unique mutations generated.

\begin{table}[t]
\centering
\caption{Dataset overview. \emph{$\pm$ Lines} shows lines added and deleted in implementation files of the \pr. \emph{Seeds/Mut} shows the number of seed test cases extracted from the \pr's InstCombine test files and the mutations generated from them.}
\label{tab:dataset}
\resizebox{\columnwidth}{!}{%
\begin{tabular}{lr r@{ /}r r@{ /}r r}
\toprule
\textbf{\pr} & \textbf{Date} & \multicolumn{2}{c}{\textbf{$\pm$ Lines}} & \multicolumn{2}{c}{\textbf{Seeds/Mut}} & \textbf{Issues} \\
\midrule
\llvmPRURL{157030} & 2025-09 & +21 & \phantom{-0} & 10 & 940 & \llvmIssueURL{156898} \\
\llvmPRURL{158097} & 2025-09 & +29 & \phantom{-0} & 16 & 1,000 & \llvmIssueURL{157371} \\
\llvmPRURL{158498} & 2025-09 & +40 & -59 & 5 & 1,000 & \llvmIssueURL{59555}, \llvmIssueURL{158326} \\
\llvmPRURL{161020} & 2025-09 & +12 & \phantom{-0} & 15 & 1,000 & \llvmIssueURL{160066} \\
\llvmPRURL{161303} & 2025-10 & +65 & -1 & 21 & 1,000 & \llvmIssueURL{157315} \\
\llvmPRURL{163628} & 2025-10 & +2 & -2 & 2 & 861 & \llvmIssueURL{162451} \\
\llvmPRURL{164733} & 2025-10 & +12 & \phantom{-0} & 9 & 1,000 & \llvmIssueURL{164436} \\
\llvmPRURL{166816} & 2025-11 & +3 & \phantom{-0} & 6 & 1,000 & \llvmIssueURL{86176}, \llvmIssueURL{163108} \\
\llvmPRURL{169960} & 2025-12 & +17 & -10 & 6 & 1,000 & \llvmIssueURL{166973} \\
\llvmPRURL{170439} & 2026-01 & +41 & \phantom{-0} & 13 & 1,000 & \llvmIssueURL{167014} \\
\llvmPRURL{171195} & 2026-01 & +11 & -2 & 5 & 1,000 & \llvmIssueURL{170020} \\
\llvmPRURL{172723} & 2026-01 & +1 & -1 & 10 & 1,000 & \llvmIssueURL{172176} \\
\llvmPRURL{173511} & 2026-01 & +58 & \phantom{-0} & 53 & 1,000 & \llvmIssueURL{154246} \\
\llvmPRURL{173768} & 2026-01 & +16 & -8 & 11 & 872 & \llvmIssueURL{173691} \\
\llvmPRURL{175876} & 2026-01 & +13 & \phantom{-0} & 5 & 1,000 & \llvmIssueURL{167178} \\
\llvmPRURL{176168} & 2026-01 & +18 & -2 & 9 & 1,000 & \llvmIssueURL{175282} \\
\llvmPRURL{177410} & 2026-01 & +36 & -12 & 16 & 1,000 & \llvmIssueURL{82350} \\
\llvmPRURL{178002} & 2026-02 & +49 & \phantom{-0} & 12 & 1,000 & \llvmIssueURL{140917} \\
\llvmPRURL{178977} & 2026-02 & +2 & -5 & 46 & 1,000 & \llvmIssueURL{172888} \\
\midrule
\textbf{Total} & 19 PRs & +446 & -102 & 270 & 18,673 & 21 \\
\bottomrule
\end{tabular}}
\end{table}


\subsubsection{Agents and Configurations}
\cref{tab:agents} summarizes the evaluated agents, models, and knowledge cutoffs~\cite{claudeModelsOverview,googleModelsGemini,openaiModelsOpenAI}.
We use default CLI configurations with a 2-hour task timeout.
Tasks failing to produce a non-empty patch are retried.
To account for stochasticity, we run each task three times per agent-model pair.

\begin{table}[htb]
  \caption{Agents, models, and knowledge cutoff dates.}
  \label{tab:agents}
  \footnotesize
  \setlength{\tabcolsep}{3pt}
  \begin{tabularx}{\columnwidth}{lXl}
    \toprule
    \textbf{Agent Version} & \textbf{Model (ID)} & \textbf{Cutoff} \\
    \midrule
    \claude 2.1.136 & \sonnetFourPointSix (claude-sonnet-4-6) & 2025-08 \\
    \gemini 0.41.2 & \geminiThreeFlash (gemini-3-flash-preview) & 2025-01 \\
    \multirow{2}{*}{\codex 0.129.0} & \gptFivePointFour (gpt-5.4) & 2025-08 \\
                            & \gptFivePointFourMini (gpt-5.4-mini) & 2025-08 \\
    \bottomrule
  \end{tabularx}
\end{table}

\subsection{RQ1: Success Rate of Agent Patches} \label{subsec:rq1}

We first build each agent patch and run seed tests via \llvmlit.
For behavioral validation on mutation-based benchmarks,
we use a two-stage pipeline:
we establish a baseline by optimizing each test case with the human patch,
retaining only cases where the resulting rewrite is \emph{behaviorally valid} (\cref{def:behavioral-validity}),
and then evaluate agent patches against these retained cases.

\subsubsection{Metrics} \label{subsubsec:metrics}
For each agent and model, we report the number of independent runs (out of three) that successfully produce patches across four metrics:
\begin{itemize}[leftmargin=*]
    \item \textbf{Buildability (\metricB)}: The patch can be applied and the \llvm project builds without errors.
    \item \textbf{Seed-test passing (\metricS)}: The patch passes all seed regression tests using \llvmlit.
    \item \textbf{Behavioral validity (\metricV)}: All mutation-based test cases result in
    behaviorally valid rewrites, ensuring no semantic or profitability regressions
    (relative to the unoptimized input) are introduced.
    \item \textbf{Non-regressive profitability (\metricP)}: All mutations are behaviorally valid
    and additionally non-regressive relative to the \emph{human-patch output},
    a stricter baseline than the unoptimized input used in \cref{def:behavioral-validity}.
    This metric reflects the primary objective of matching the optimization performance
    of human-written patches.
\end{itemize}

We also report Pass@1, the fraction of all runs (three per \pr) that
achieve the metric, and Pass@3, the fraction of \prs for which at least
one of the three runs achieves it.

\subsubsection{Results}

\begin{table}[t]
  \caption{Success counts per \pr across three independent runs for each agent and model configuration.
  Each cell reports four metrics (\metricB/\metricS/\metricV/\metricP) as explained in \cref{subsubsec:metrics};
  darker green indicates more successful runs (0--3).
  }
  \label{tab:rq1}
  \footnotesize
  \resizebox{\columnwidth}{!}{%
  \begin{tabular}{l *{4}{w{c}{0.4em}}| *{4}{w{c}{0.4em}}| *{4}{w{c}{0.4em}}| *{4}{w{c}{0.4em}}}
    \toprule
    & \multicolumn{4}{c}{\claude} & \multicolumn{4}{c}{\gemini} & \multicolumn{8}{c}{\codex} \\
    \cmidrule(lr){2-5}\cmidrule(lr){6-9}\cmidrule(lr){10-17}
    \textbf{19 \prs} & \multicolumn{4}{c}{\scriptsize{\sonnetFourPointSix}} & \multicolumn{4}{c}{\scriptsize{\geminiThreeFlash}} & \multicolumn{4}{c}{\scriptsize{\gptFivePointFour}} & \multicolumn{4}{c}{\scriptsize{\gptFivePointFourMini}} \\
    \midrule
    & \metricB & \metricS & \metricV & \metricP & \metricB & \metricS & \metricV & \metricP & \metricB & \metricS & \metricV & \metricP & \metricB & \metricS & \metricV & \metricP \\
    \midrule
    \llvmPRURL{157030} & \cellcolor{g3}3 & \cellcolor{white}0 & \cellcolor{white}0 & \cellcolor{white}0 & \cellcolor{g3}3 & \cellcolor{white}0 & \cellcolor{g1}1 & \cellcolor{white}0 & \cellcolor{g3}3 & \cellcolor{white}0 & \cellcolor{g1}1 & \cellcolor{white}0 & \cellcolor{g3}3 & \cellcolor{white}0 & \cellcolor{white}0 & \cellcolor{white}0 \\
    \llvmPRURL{158097} & \cellcolor{g3}3 & \cellcolor{g3}3 & \cellcolor{g3}3 & \cellcolor{white}0 & \cellcolor{g3}3 & \cellcolor{g1}1 & \cellcolor{g1}1 & \cellcolor{g1}1 & \cellcolor{g3}3 & \cellcolor{g2}2 & \cellcolor{g2}2 & \cellcolor{g2}2 & \cellcolor{g3}3 & \cellcolor{g1}1 & \cellcolor{g3}3 & \cellcolor{white}0 \\
    \llvmPRURL{158498} & \cellcolor{g3}3 & \cellcolor{white}0 & \cellcolor{g3}3 & \cellcolor{white}0 & \cellcolor{g3}3 & \cellcolor{g3}3 & \cellcolor{g3}3 & \cellcolor{g3}3 & \cellcolor{g3}3 & \cellcolor{g2}2 & \cellcolor{g2}2 & \cellcolor{g2}2 & \cellcolor{g2}2 & \cellcolor{white}0 & \cellcolor{white}0 & \cellcolor{white}0 \\
    \llvmPRURL{161020} & \cellcolor{g3}3 & \cellcolor{g1}1 & \cellcolor{g3}3 & \cellcolor{g2}2 & \cellcolor{g3}3 & \cellcolor{white}0 & \cellcolor{g2}2 & \cellcolor{g2}2 & \cellcolor{g3}3 & \cellcolor{white}0 & \cellcolor{g1}1 & \cellcolor{g1}1 & \cellcolor{g3}3 & \cellcolor{white}0 & \cellcolor{g2}2 & \cellcolor{g2}2 \\
    \llvmPRURL{161303} & \cellcolor{g3}3 & \cellcolor{white}0 & \cellcolor{g3}3 & \cellcolor{white}0 & \cellcolor{g3}3 & \cellcolor{g1}1 & \cellcolor{g2}2 & \cellcolor{white}0 & \cellcolor{g3}3 & \cellcolor{g1}1 & \cellcolor{g3}3 & \cellcolor{white}0 & \cellcolor{g3}3 & \cellcolor{white}0 & \cellcolor{g1}1 & \cellcolor{white}0 \\
    \llvmPRURL{163628} & \cellcolor{g3}3 & \cellcolor{g3}3 & \cellcolor{g3}3 & \cellcolor{g3}3 & \cellcolor{g3}3 & \cellcolor{g2}2 & \cellcolor{g3}3 & \cellcolor{g3}3 & \cellcolor{g3}3 & \cellcolor{g2}2 & \cellcolor{g3}3 & \cellcolor{g3}3 & \cellcolor{g3}3 & \cellcolor{g3}3 & \cellcolor{g3}3 & \cellcolor{g3}3 \\
    \llvmPRURL{164733} & \cellcolor{g3}3 & \cellcolor{g1}1 & \cellcolor{g3}3 & \cellcolor{g1}1 & \cellcolor{g3}3 & \cellcolor{white}0 & \cellcolor{g3}3 & \cellcolor{white}0 & \cellcolor{g3}3 & \cellcolor{white}0 & \cellcolor{g3}3 & \cellcolor{white}0 & \cellcolor{g3}3 & \cellcolor{white}0 & \cellcolor{g3}3 & \cellcolor{white}0 \\
    \llvmPRURL{166816} & \cellcolor{g3}3 & \cellcolor{white}0 & \cellcolor{white}0 & \cellcolor{white}0 & \cellcolor{g3}3 & \cellcolor{g2}2 & \cellcolor{white}0 & \cellcolor{white}0 & \cellcolor{g3}3 & \cellcolor{white}0 & \cellcolor{white}0 & \cellcolor{white}0 & \cellcolor{g3}3 & \cellcolor{g1}1 & \cellcolor{white}0 & \cellcolor{white}0 \\
    \llvmPRURL{169960} & \cellcolor{g3}3 & \cellcolor{g2}2 & \cellcolor{g3}3 & \cellcolor{g3}3 & \cellcolor{g3}3 & \cellcolor{g2}2 & \cellcolor{white}0 & \cellcolor{white}0 & \cellcolor{g3}3 & \cellcolor{white}0 & \cellcolor{g3}3 & \cellcolor{white}0 & \cellcolor{g3}3 & \cellcolor{g1}1 & \cellcolor{g2}2 & \cellcolor{g2}2 \\
    \llvmPRURL{170439} & \cellcolor{g3}3 & \cellcolor{white}0 & \cellcolor{white}0 & \cellcolor{white}0 & \cellcolor{g3}3 & \cellcolor{white}0 & \cellcolor{white}0 & \cellcolor{white}0 & \cellcolor{g3}3 & \cellcolor{white}0 & \cellcolor{white}0 & \cellcolor{white}0 & \cellcolor{g3}3 & \cellcolor{white}0 & \cellcolor{white}0 & \cellcolor{white}0 \\
    \llvmPRURL{171195} & \cellcolor{g3}3 & \cellcolor{white}0 & \cellcolor{white}0 & \cellcolor{white}0 & \cellcolor{g3}3 & \cellcolor{white}0 & \cellcolor{white}0 & \cellcolor{white}0 & \cellcolor{g3}3 & \cellcolor{white}0 & \cellcolor{white}0 & \cellcolor{white}0 & \cellcolor{g3}3 & \cellcolor{white}0 & \cellcolor{white}0 & \cellcolor{white}0 \\
    \llvmPRURL{172723} & \cellcolor{g3}3 & \cellcolor{white}0 & \cellcolor{g3}3 & \cellcolor{white}0 & \cellcolor{g3}3 & \cellcolor{g1}1 & \cellcolor{g1}1 & \cellcolor{g1}1 & \cellcolor{g3}3 & \cellcolor{white}0 & \cellcolor{g3}3 & \cellcolor{white}0 & \cellcolor{g3}3 & \cellcolor{white}0 & \cellcolor{g3}3 & \cellcolor{white}0 \\
    \llvmPRURL{173511} & \cellcolor{g3}3 & \cellcolor{white}0 & \cellcolor{white}0 & \cellcolor{white}0 & \cellcolor{g3}3 & \cellcolor{white}0 & \cellcolor{white}0 & \cellcolor{white}0 & \cellcolor{g3}3 & \cellcolor{white}0 & \cellcolor{white}0 & \cellcolor{white}0 & \cellcolor{g3}3 & \cellcolor{white}0 & \cellcolor{g1}1 & \cellcolor{white}0 \\
    \llvmPRURL{173768} & \cellcolor{g3}3 & \cellcolor{g1}1 & \cellcolor{g3}3 & \cellcolor{g1}1 & \cellcolor{g3}3 & \cellcolor{g1}1 & \cellcolor{g3}3 & \cellcolor{g2}2 & \cellcolor{g3}3 & \cellcolor{white}0 & \cellcolor{g3}3 & \cellcolor{white}0 & \cellcolor{g3}3 & \cellcolor{white}0 & \cellcolor{g2}2 & \cellcolor{white}0 \\
    \llvmPRURL{175876} & \cellcolor{g3}3 & \cellcolor{g3}3 & \cellcolor{g3}3 & \cellcolor{g3}3 & \cellcolor{g3}3 & \cellcolor{g3}3 & \cellcolor{g2}2 & \cellcolor{g2}2 & \cellcolor{g3}3 & \cellcolor{g2}2 & \cellcolor{g3}3 & \cellcolor{g2}2 & \cellcolor{g3}3 & \cellcolor{g1}1 & \cellcolor{g3}3 & \cellcolor{g1}1 \\
    \llvmPRURL{176168} & \cellcolor{g3}3 & \cellcolor{white}0 & \cellcolor{g3}3 & \cellcolor{white}0 & \cellcolor{g3}3 & \cellcolor{white}0 & \cellcolor{white}0 & \cellcolor{white}0 & \cellcolor{g3}3 & \cellcolor{white}0 & \cellcolor{g2}2 & \cellcolor{white}0 & \cellcolor{g3}3 & \cellcolor{white}0 & \cellcolor{g2}2 & \cellcolor{white}0 \\
    \llvmPRURL{177410} & \cellcolor{g3}3 & \cellcolor{white}0 & \cellcolor{g3}3 & \cellcolor{g1}1 & \cellcolor{g3}3 & \cellcolor{white}0 & \cellcolor{g3}3 & \cellcolor{g2}2 & \cellcolor{g3}3 & \cellcolor{white}0 & \cellcolor{g1}1 & \cellcolor{g1}1 & \cellcolor{g3}3 & \cellcolor{white}0 & \cellcolor{white}0 & \cellcolor{white}0 \\
    \llvmPRURL{178002} & \cellcolor{g3}3 & \cellcolor{white}0 & \cellcolor{g3}3 & \cellcolor{white}0 & \cellcolor{g3}3 & \cellcolor{white}0 & \cellcolor{g3}3 & \cellcolor{white}0 & \cellcolor{g3}3 & \cellcolor{white}0 & \cellcolor{g3}3 & \cellcolor{white}0 & \cellcolor{g3}3 & \cellcolor{white}0 & \cellcolor{g3}3 & \cellcolor{white}0 \\
    \llvmPRURL{178977} & \cellcolor{g3}3 & \cellcolor{white}0 & \cellcolor{white}0 & \cellcolor{white}0 & \cellcolor{g3}3 & \cellcolor{white}0 & \cellcolor{white}0 & \cellcolor{white}0 & \cellcolor{g3}3 & \cellcolor{g2}2 & \cellcolor{g3}3 & \cellcolor{g2}2 & \cellcolor{g3}3 & \cellcolor{g3}3 & \cellcolor{g3}3 & \cellcolor{g3}3 \\
    \midrule
    \textbf{\scriptsize Pass@3 (\%)} & \textbf{100} & 37 & \evalRqIValidityPassAtThreeOfSonnetFourPointSixAndGptFivePointFourMini{} & 37 & \textbf{100} & \textbf{47} & \evalRqIValidityPassAtThreeOfGeminiThreeFlash{} & \textbf{\evalRqIProfitabilityPassAtThreeOfGeminiThreeFlash{}} & \textbf{100} & 32 & \textbf{\evalRqIValidityPassAtThreeOfGptFivePointFour{}} & 37 & \textbf{100} & 32 & \evalRqIValidityPassAtThreeOfSonnetFourPointSixAndGptFivePointFourMini{} & 26 \\
    \textbf{\scriptsize Pass@1 (\%)} & \textbf{100} & 25 & \textbf{68} & 25 & \textbf{100} & \textbf{28} & 47 & \textbf{\evalRqIProfitabilityPassAtOneOfGeminiThreeFlash{}} & \textbf{100} & 19 & 63 & 23 & 98 & 18 & 54 & 19 \\
    \bottomrule
  \end{tabular}}
\end{table}

\cref{tab:rq1} summarizes the results.
Models show near-perfect buildability (100\% Pass@3),
indicating their proficiency in navigating the \llvm codebase and iteratively resolving diagnostics.
The only build failures stem from an agent skipping validation and
submitting an uncompiled patch with undeclared identifiers.

\myfinding[finding:rq1.1]{
Coding agents achieve near-perfect buildability,
though they occasionally submit patches without validation.
}

Behavioral validity (\metricV) is the minimum safety bar against regressions.
\codex models lead here,
with \gptFivePointFour achieving \evalRqIValidityPassAtThreeOfGptFivePointFour{\%} Pass@3,
followed by \gptFivePointFourMini (\evalRqIValidityPassAtThreeOfSonnetFourPointSixAndGptFivePointFourMini{\%}).
\geminiThreeFlash ranks lowest (\evalRqIValidityPassAtThreeOfGeminiThreeFlash{\%}).

\myfinding[finding:rq1.2]{
\codex models reliably maintain behavioral validity (up to \evalRqIValidityPassAtThreeOfGptFivePointFour{\%}),
while \geminiThreeFlash achieves the lowest (\evalRqIValidityPassAtThreeOfGeminiThreeFlash{\%}).
}

However, achieving behavioral validity does not guarantee
meeting the stricter non-regressive profitability objective relative to the human patch.
\geminiThreeFlash leads in profitability with
\evalRqIProfitabilityPassAtThreeOfGeminiThreeFlash{\%} Pass@3
and \evalRqIProfitabilityPassAtOneOfGeminiThreeFlash{\%} Pass@1.

\myfinding[finding:rq1.3]{
\geminiThreeFlash leads the stricter profitability objective
(\evalRqIProfitabilityPassAtThreeOfGeminiThreeFlash{\%} Pass@3),
yet matching human experts across both metrics remains challenging.
}

Comparing seed-test passing and profitability
reveals two limitations of traditional test suites.
First, cases where patches achieve profitability 
but fail seed tests ($\metricP > \metricS$, \eg, \llvmPRURL{161020}) indicate that
\llvm's exact \filecheck patterns are too brittle,
rejecting valid, profitable optimizations.
Conversely, cases where patches pass seed tests 
but regress in profitability 
($\metricS$$ > $$\metricP$, \eg, \llvmPRURL{158097} for \sonnetFourPointSix) 
demonstrate that seed tests lack the coverage to 
expose subtle performance regressions.

\myfinding[finding:rq1.4]{
Discrepancies between seed-test passing and 
profitability highlight dual test suite limitations: 
brittle \filecheck patterns reject valid optimizations ($\metricP > \metricS$), 
while limited coverage of edge cases in seed tests
masks profitability regressions ($\metricS > \metricP$).
}

\subsubsection{Case Studies}
\label{subsubsec:rq1:case-studies}

We illustrate the dual limitations of seed test suites identified in \cref{finding:rq1.4}
with two representative PRs.

\paragraph{\filecheck rejects profitable optimizations ($\metricP > \metricS$)} \label{par:rq1:case-studies:fragile}
\llvmPRURL{161020} implements a fold that rewrites
the sum-reduction of a splat vector (a vector of identical elements)
to a scalar multiplication:
\code{$\text{reduce.add(splat(x, N))} \;\to\; \text{mul x, N}$}.
As shown in \cref{fig:rq1:cs:161020:input}, the seed test uses an \code{i1}
element type with $N = 8$.
In the human patch, an existing fold that handles \code{i1} vectors
by rewriting \code{reduce.add} to a \code{ctpop} (count the number of 1-bits)
sequence takes priority over the newly added fold,
so the \code{CHECK} directives in the seed test expect \code{ctpop}.
The agent, however, gives the new fold higher priority so that it fires first.
For an \code{i1} vector of size~8, \code{mul i1 \%x, 8} constant-folds
to 0 (since $8 \bmod 2 = 0$),
yielding the output shown in \cref{fig:rq1:cs:161020:agent}.
This optimization is semantically correct and more profitable,
yet the rigid \filecheck patterns cause the seed tests to reject it.
We have submitted a \pr (\anonymizedLlvmPRURL{182213}{A74E})
to address this issue by adjusting the fold priority,
and it has been merged.

\begin{figure}[t]
    \begin{subfigure}[t]{\columnwidth}
        \TSCodeInput{figures/rq1-cs-161020-input.ts.tex}
        \caption{A seed test case with an \code{i1} splat vector of size 8.
        The input \ir constructs the splat via \code{insertelement} and
        \code{shufflevector}, then reduces by addition.
        The \filecheck directives expect the human patch to rewrite
        \code{reduce.add} into a \code{ctpop}-based sequence.}
        \label{fig:rq1:cs:161020:input}
        \vspace{0.5em}
    \end{subfigure}
    \begin{subfigure}[t]{\columnwidth}
        \TSCodeInput{figures/rq1-cs-161020-agent.ts.tex}
        \caption{The optimized \ir produced by agent patches.
        By applying the splat-to-multiplication fold with higher priority,
        the agent yields \code{mul i1 \%x, 8}, which folds to \code{0} and is more profitable.}
        \label{fig:rq1:cs:161020:agent}
    \end{subfigure}
    \caption{Rigid \filecheck patterns in \llvmPRURL{161020}
    reject a correct and more profitable optimization produced by the agent.}
    \label{fig:rq1:cs:161020}
\end{figure}

\paragraph{Seed tests mask profitability regressions ($\metricS > \metricP$)}
\label{par:rq1:case-studies:coverage}
\llvmPRURL{158097} folds a disjunction of two \code{fcmp} (floating-point comparison)
instructions sharing the same variable and predicate into a single \code{fcmp}.
As shown in \cref{fig:rq1:cs:158097:input}, the mutation applies \code{@llvm.fake.use}---a
test intrinsic that simulates a downstream use, keeping a value live---to the
second \code{fcmp} (\code{\%v2}).
Since \code{\%v2} is exactly the single \code{fcmp} the fold would produce,
the human patch simply reuses it as the return value without introducing new instructions.
\sonnetFourPointSix instead introduces a redundant fresh \code{fcmp} (\cref{fig:rq1:cs:158097:agent}),
a regression the seed tests never expose because they contain no
\code{@llvm.fake.use} cases on \code{fcmp} operands.

\begin{figure}[t]
    \begin{subfigure}[t]{\columnwidth}
        \TSCodeInput{figures/rq1-cs-158097-input.ts.tex}
        \caption{A mutation computing $v_0 < 0.01 \lor v_0 < 1.99$,
        where \code{\%v2} is consumed by \code{@llvm.fake.use} to
        simulate a downstream use.}
        \label{fig:rq1:cs:158097:input}
        \vspace{0.5em}
    \end{subfigure}
    \begin{subfigure}[t]{\columnwidth}
        \TSCodeInput{figures/rq1-cs-158097-agent.ts.tex}
        \caption{The optimized \ir produced by the agent.
        Instead of reusing \code{\%v2}, the agent introduces a fresh
        \code{\%v3} with identical operands and predicate,
        leaving two \code{fcmp} instructions rather than one.}
        \label{fig:rq1:cs:158097:agent}
    \end{subfigure}
    \caption{Seed tests in \llvmPRURL{158097} lack coverage of
    \code{@use}, masking a profitability regression
    where the agent introduces a redundant \code{fcmp} instead of reusing
    the existing one.}
    \label{fig:rq1:cs:158097}
\end{figure}

\subsection{RQ2: Behavioral Comparison Against Human Patches} \label{subsec:rq2}

In RQ2, we directly compare agent and human patches on the mutation-based benchmarks.
We retain test cases on which the \emph{unpatched} compiler's output is
behaviorally valid (unlike RQ1, we do not filter on human-patch validity),
and exclude test cases for which \alivetwo is inconclusive for the human or any agent rewrite.
We report \textbf{behavioral validity} (\cref{def:behavioral-validity}),
\textbf{profitability} (percentage of valid rewrites that improve over the base commit per \cref{def:profitability}),
and \textbf{generalization} (percentage of test cases where the agent produces a profitable rewrite but the human patch does not).
Note that profitability here is measured relative to the \emph{unpatched base commit},
a weaker criterion than RQ1's metric (\metricP), which requires agents to match the human patch output.

\subsubsection{Results}

The aggregate results for RQ2 are presented in \cref{tab:rq2-aggregate}.
Across all agents, we observe high behavioral validity (\evalRqIIAgentValidityLowest{\%}--\evalRqIIAgentValidityHighest{\%}),
which is comparable to or even exceeds the human-written patches (\evalRqIIHumanValidity{\%}).
The fact that human patches do not achieve 100\% validity is notable;
we categorize human patch failures into three buckets,
summarized in \cref{tab:rq2-human-validity}.

\begin{table}[t]
  \centering
  \caption{Percentage of test cases that are valid, profitable, and
  generalized-profitable relative to the human patch,
  averaged over three runs per \pr and then across \prs with equal weight.
  The values support relative comparison under the same generated test
  distribution rather than estimates of absolute optimization performance.}
  \label{tab:rq2-aggregate}
  \footnotesize
  \begin{tabular}{llrrr}
    \toprule
    \textbf{Agent} & \textbf{Model} & \textbf{Valid} & \textbf{Profitable} & \textbf{Generalized} \\
    \midrule
    Human & Pull Request & \evalRqIIHumanValidity{\%} & \evalRqIIHumanProfitability{\%} & - \\
    Claude & Claude Sonnet 4.6 & \textbf{\evalRqIIAgentValidityHighest{\%}} & 9.0\% & \textbf{0.7\%} \\
    Gemini & Gemini 3 Flash & 97.7\% & \textbf{\evalRqIIGeminiProfitability{\%}} & \textbf{\evalRqIIGeminiGeneralization{\%}} \\
    Codex & GPT-5.4 & \textbf{\evalRqIIAgentValidityLowest{\%}} & 9.0\% & \textbf{0.7\%} \\
    Codex & GPT-5.4 Mini & 97.5\% & \textbf{\evalRqIIOtherAgentsProfitabilityLow{\%}} & 1.1\% \\
    \bottomrule
  \end{tabular}
\end{table}

The dominant category is \emph{profitability regression} (\evalRqIIHumanProfitabilityIssueInvalidCount{} of \evalRqIIHumanInvalidTotal{} cases,
\evalRqIIHumanProfitabilityIssueInvalidPct{\%}):
the human patch produces a rewrite that is regressive relative to the base commit.
Among these, about \evalRqIIHumanLargeRegressionPct{\%} of the \prs have cases with a cycle regression $\geq$ 100,
indicating that the regressions are not minor.
A single failure is due to \emph{optimization failure} (unreachable fixpoint),
where the \instcombine pass failed to converge.
Notably, no cases of \emph{incorrect transformation} were identified,
confirming that human patches generally maintain semantic correctness.

\myfinding[finding:rq2.1]{
Human-written patches are not always behaviorally valid,
but failures are almost exclusively profitability regressions rather than semantic incorrectness.
}

In terms of validity, \sonnetFourPointSix achieves the highest rate (\evalRqIIAgentValidityHighest{\%}),
matching the human-written baseline (\evalRqIIHumanValidity{\%}),
while \geminiThreeFlash lags slightly behind at \evalRqIIGeminiValidity{\%}.
In terms of profitability, however,
\geminiThreeFlash achieves \evalRqIIGeminiProfitability{\%}, on par with the human-written baseline (\evalRqIIHumanProfitability{\%}),
whereas all other agents fall well short at \evalRqIIOtherAgentsProfitabilityLow{\%}--\evalRqIIOtherAgentsProfitabilityHigh{\%}.
\geminiThreeFlash also leads in generalization (\evalRqIIGeminiGeneralization{\%}),
discovering profitable optimizations in cases that the human patch failed to cover.

\myfinding[finding:rq2.2]{
Agents reveal a trade-off between validity and profitability:
\sonnetFourPointSix achieves the highest validity matching human experts
but well below human profitability,
while \geminiThreeFlash matches human profitability
and leads in generalization at a small validity cost.
}

\begin{table}[t]
  \centering
  \caption{Breakdown of human-patch testcases classified as behaviorally invalid across \numPRs \prs, by failure reason.
  The \prs / \pr\% columns report how many of the \numPRs \prs contribute at least one such testcase in any agent's run.}
  \label{tab:rq2-human-validity}
  \footnotesize
  \begin{tabular}{lrrrr}
    \toprule
    \textbf{Failure Reason} & \textbf{Count} & \textbf{\%} & \textbf{\prs} & \textbf{\pr\%} \\
    \midrule
    Profitability regression & 287 & 99.7\% & 9 & 47.4\% \\
    \quad $\Delta$cycles $\geq$ 100 & 21 & 7.3\% & 3 & 15.8\% \\
    \addlinespace
    Optimization failure & 1 & 0.3\% & 1 & 5.3\% \\
    Incorrect transformation & 0 & 0.0\% & 0 & 0.0\% \\
    \midrule
    \textbf{Total} & \textbf{288} & \textbf{100\%} & \textbf{9} & \textbf{47.4\%} \\
    \bottomrule
  \end{tabular}
\end{table}

\subsubsection{Case Studies}

\paragraph{Human patch regression in the multi-use case}
\llvmPRURL{157030} folds $\mathrm{trunc}_{\mathrm{i1}}(C \gg x)$ to $x \mathbin{==} \log_2 C$
for power-of-two constants~$C$, exploiting the fact that $C \gg x$ yields~1
only when $x$ equals the bit position of~$C$.
The fold is profitable only when the result of \code{lshr} becomes dead and
can be eliminated.
As shown in \cref{fig:rq2:cs:157030}, the test case includes an
\code{@llvm.fake.use} call that creates a second use of \code{\%lshr},
preventing its elimination.
When the fold fires, it introduces \code{icmp eq i16 \%x,~4}
(since $16 = 2^4$) but leaves \code{\%lshr} in place,
yielding an output that carries the original \code{lshr} alongside the new
comparison. As \code{icmp} is more expensive than \code{trunc},
this rewrite is regressive.

\begin{figure}[ht]
    \TSCodeInput{figures/rq2-cs-157030-input.ts.tex}
    \caption{A test case for \llvmPRURL{157030} where \code{@llvm.fake.use}
    creates a second use of \code{\%lshr}.
    The fold fires but cannot eliminate \code{\%lshr},
    leaving it live alongside the new \code{icmp eq}
    and causing a profitability regression.}
    \label{fig:rq2:cs:157030}
\end{figure}

\paragraph{Agent generalization beyond the human patch}
We revisit \llvmPRURL{158097} from the generalization angle
(see \cref{par:rq1:case-studies:coverage} for the test coverage perspective).
The human patch implements the redundant-\code{fcmp} fold only for ordered
predicates (\eg, \code{olt}), leaving the symmetric unordered forms
(\eg, \code{ult}) without a corresponding rule.
Several agent patches further extend the fold to unordered predicates.

\subsection{RQ3: Failure Modes of Agent Patches} \label{subsec:rq3}

Following the methodology in \cref{subsec:failure-mode-analysis},
we organize failure modes into ten subcategories under three high-level categories
(\cref{tab:rq3-agent-failures}): profitability issues, optimization failures, and incorrect transformations.
We construct a representative set of \numLabelCases failure cases via stratified sampling;
Fleiss's $\kappa = \kappaScore$ across the three labelers indicates almost perfect agreement~\cite{landis1977measurement,fleiss1971measuring}.

\begin{table}[t]
  \centering
  \caption{Failure-mode breakdown of unsuccessful agent patches across \numPRs PRs.
  Values count the number of distinct PRs contributing at least one case in any agent's run.}
  \label{tab:rq3-agent-failures}
  \footnotesize
  \setlength{\tabcolsep}{2pt}
  \resizebox{\columnwidth}{!}{%
  \begin{tabular}{p{3.2cm}cccc}
    \toprule
    \textbf{Failure Mode} & \scriptsize{\sonnetFourPointSix} & \scriptsize{\geminiThreeFlash} & \scriptsize{\gptFivePointFour} & \scriptsize{\gptFivePointFourMini} \\
    \midrule
    \raggedright \textbf{Profitability issue} & \textbf{16} & \textbf{16} & \textbf{15} & \textbf{16} \\
    \raggedright \quad No change & 14 & 9 & 15 & 15 \\
    \raggedright \quad Regressive \vs\ input \\ \quad \textit{(slower than no change)} & 5 & 7 & 4 & 8 \\
    \raggedright \quad Profitable \vs\ baseline \\ \quad \textit{(but below human patch)} & 3 & 5 & 2 & 4 \\
    \raggedright \quad Mixed \vs\ baseline & 0 & 1 & 0 & 0 \\
    \addlinespace
    \raggedright \textbf{Optimization failure} & \textbf{1} & \textbf{3} & \textbf{3} & \textbf{2} \\
    \raggedright \quad Unreachable fixpoint & 0 & 2 & 1 & 0 \\
    \raggedright \quad Timeout & 1 & 2 & 2 & 1 \\
    \raggedright \quad Mismatched return type & 0 & 0 & 1 & 1 \\
    \addlinespace
    \raggedright \textbf{Incorrect transformation} & \textbf{0} & \textbf{4} & \textbf{5} & \textbf{7} \\
    \raggedright \quad With \code{undef} input & 0 & 1 & 1 & 0 \\
    \raggedright \quad Target is more poisonous & 0 & 2 & 3 & 2 \\
    \raggedright \quad Other (value mismatch) & 0 & 1 & 3 & 5 \\
    \midrule
    \raggedright \textbf{Total \prs} & \textbf{16} & \textbf{17} & \textbf{17} & \textbf{17} \\
    \bottomrule
  \end{tabular}}
\end{table}

\subsubsection{Results and Case Studies}

\cref{tab:rq3-themes} summarizes the \numThemes failure themes
under three observed high-level failure modes.
We discuss each in turn.

\begin{table}[t]
  \centering
  \caption{Failure themes of unsuccessful agent patches derived
  from a qualitative coding of \numLabelCases cases
  (131 with unanimous agreement, 29 resolved by human adjudication),
  broken down by the same failure-mode categories as Table~\ref{tab:rq3-agent-failures}.
  Values count the number of distinct PRs (of \numPRs) contributing at least one case of that theme in any agent's run.}
  \label{tab:rq3-themes}
  \footnotesize
  \setlength{\tabcolsep}{3pt}
  \resizebox{\columnwidth}{!}{%
  \begin{tabular}{p{3cm} c c c c}
    \toprule
    \textbf{Theme / Failure Mode} & \scriptsize{\sonnetFourPointSix} & \scriptsize{\geminiThreeFlash} & \scriptsize{\gptFivePointFour} & \scriptsize{\gptFivePointFourMini} \\
    \midrule
    \raggedright \textbf{Under-generalization} & \textbf{13} & \textbf{9} & \textbf{13} & \textbf{13} \\
    \raggedright \textbf{Canonicalization issues} & \textbf{3} & \textbf{3} & \textbf{2} & \textbf{3} \\
    \raggedright \textbf{Fold interaction} & \textbf{2} & \textbf{2} & \textbf{1} & \textbf{2} \\
    \multicolumn{5}{@{}l@{}}{\scriptsize\itshape \quad The three themes above occur under \emph{profitability issues} only.} \\
    \addlinespace
    \raggedright \textbf{Incorrect one-use guard} & \textbf{4} & \textbf{5} & \textbf{4} & \textbf{6} \\
    \raggedright \; Profitability issues & 4 & 5 & 4 & 6 \\
    \raggedright \; Optimization failure & 0 & 0 & 1 & 0 \\
    \addlinespace
    \raggedright \textbf{Poison/FMF flag misuse} & \textbf{0} & \textbf{4} & \textbf{3} & \textbf{1} \\
    \raggedright \; Profitability issues & 0 & 2 & 1 & 1 \\
    \raggedright \; Incorrect transformation & 0 & 3 & 2 & 0 \\
    \addlinespace
    \raggedright \textbf{Unsound transform} & \textbf{0} & \textbf{1} & \textbf{5} & \textbf{7} \\
    \raggedright \; Optimization failure & 0 & 0 & 1 & 1 \\
    \raggedright \; Incorrect transformation & 0 & 1 & 4 & 7 \\
    \addlinespace
    \raggedright \textbf{Non-convergent rewrite} & \textbf{1} & \textbf{3} & \textbf{2} & \textbf{1} \\
    \multicolumn{5}{@{}l@{}}{\scriptsize\itshape \quad The theme above occurs under \emph{optimization failure} only.} \\
    \midrule
    \raggedright \textbf{Total} & \textbf{16} & \textbf{17} & \textbf{17} & \textbf{17} \\
    \bottomrule
  \end{tabular}}
\end{table}

\paragraph{Under-Generalization} \label{par:rq3:under-generalization}

This is the most common failure theme, observed in 9--13 \prs.
It captures cases where the agent-generated patch implements a correct optimization
but with tighter preconditions or missing matching rules compared to the human patch.
The only failure mode observed under this theme is \emph{no change},
\ie, the input \ir is left unchanged by the agent patch.
The example in \cref{fig:background-instcombine-example:agent}
illustrates this failure theme,
where the agent patch misses the fold for constants that are low-bit masks.

\paragraph{Canonicalization Issue} \label{par:rq3:canonicalization}

Sometimes, the \ir produced by the agent patch has the same
instruction sequence as the human patch's output but in a different canonical form,
such as using a different operand order or comparison predicate.
These differences may trigger different lowering paths in \llvm's backend,
leading to different profitability outcomes.
For example, \Cref{fig:rq3:cs:canon} shows two semantically equivalent \ir{} functions
that differ only in the operand order of a commutative \code{mul}.
The non-canonical form in \Cref{fig:rq3:cs:canon:a} generates
an extra \code{mov} instruction in the output assembly
compared to the canonical form in \Cref{fig:rq3:cs:canon:b},
causing the agent patch's output to be regressive.
This could be a missed optimization in \llvm's backend.
Canonicalization issues are not as critical as other failure modes,
since most cases are still \emph{profitable \vs baseline}.
However, they obscure the true performance of the patch and
make it difficult to reliably evaluate profitability.

\begin{figure}[t]
    \begin{subfigure}[t]{\columnwidth}
        \TSCodeInput{figures/rq3-canon-a.ts.tex}
        \caption{Agent patch: \code{mul i8 \%y, \%1} (parameter \code{\%y} first).
        The backend loads \code{\%y} into the register via an extra \code{mov} instruction.}
        \label{fig:rq3:cs:canon:a}
        \vspace{0.5em}
    \end{subfigure}
    \begin{subfigure}[t]{\columnwidth}
        \TSCodeInput{figures/rq3-canon-b.ts.tex}
        \caption{Human patch: \code{mul i8 \%1, \%y} (\code{select} result \code{\%1} first).}
        \label{fig:rq3:cs:canon:b}
    \end{subfigure}
    \caption{Canonicalization issue: \code{mul} with swapped operands;
    the non-canonical form (top) generates one extra \code{mov}.}
    \label{fig:rq3:cs:canon}
\end{figure}

\paragraph{Fold Interaction} \label{par:rq3:fold-interaction}

In these cases, the patch places the new fold at a different position in the match chain,
producing a different fold firing order
that preemptively blocks more profitable folds from triggering.
\Cref{fig:rq3:cs:fold-prio:input} shows a representative case from \llvmPRURL{157030}.
The human patch places the new fold (\code{trunc} to \code{icmp}) 
at line 970 (\cref{fig:rq3:cs:fold-prio:human}),
so an existing fold at line 963 fires first and produces \code{true}. 
The agent inserts the new fold 
at line 848 (\cref{fig:rq3:cs:fold-prio:agent}), 
so it fires first and blocks the existing fold at line 963, 
causing a regression on profitability.

\begin{figure}[t]
    \begin{subfigure}[t]{\columnwidth}
        \TSCodeInput{figures/rq3-fold-prio-input.ts.tex}
        \caption{Input \ir: \code{trunc nsw (lshr exact i8 -128, \%x)}, where
        $-128$ is a power of two and
        \code{\%lshr} has an additional use via \code{@use}.}
        \label{fig:rq3:cs:fold-prio:input}
        \vspace{0.5em}
    \end{subfigure}
    \begin{subfigure}[t]{\columnwidth}
        \TSCodeInput{figures/rq3-fold-prio-human.ts.tex}
        \caption{Human patch: the new fold is inserted at line 970, after the existing
        fold at line 963. For the input in (a), line 963 fires first
        and folds \code{trunc} to \code{true}.}
        \label{fig:rq3:cs:fold-prio:human}
    \end{subfigure}
    \begin{subfigure}[t]{\columnwidth}
        \TSCodeInput{figures/rq3-fold-prio-agent.ts.tex}
        \caption{Agent patch: the new fold is inserted at line 848, before the existing
        fold at line 963. For the input in (a), the new fold fires first,
        blocking the more profitable fold at line 963.}
        \label{fig:rq3:cs:fold-prio:agent}
        \vspace{0.5em}
    \end{subfigure}
    \caption{Fold interaction failure on \llvmPRURL{157030}.}
    \label{fig:rq3:cs:fold-prio}
\end{figure}

\paragraph{Incorrect One-Use Guard} \label{par:rq3:one-use-guard}

Patches in this theme fail to correctly implement 
the one-use guard for a fold---missing
it entirely, adding an unnecessary guard, or confusing \code{hasOneUse} with \code{hasOneUser}
(one user may have multiple uses, \eg, a binary operation on identical operands).
This is the second most common failure theme across all agents,
observed in four to six \prs per model.
One-use guards are a pervasive and critical precondition in \instcombine:
without the guard, a fold may introduce an additional instruction, making the rewrite regressive.
\cref{fig:rq2:cs:157030} illustrates a missing one-use guard,
where the fold fires even when the \code{lshr} instruction has multiple uses,
leaving a regressive rewrite.
What is worse, they can cause non-convergence.
The \ir in \cref{fig:rq3:cs:one-use-nc} contains a chain of two
\code{getelementptr}~\cite{llvmOftenMisunderstood} instructions,
performing pointer arithmetic equivalent to \code{((i8*)y+a)+8148}.
The agent's rewrite attempts to fold
this into \code{(i8*)y+(add a,8148)} 
when \code{\%a} is a \code{select} result,
but omits a one-use guard on the \code{select}.
Because the intermediate pointer \code{\%g1} has an additional use (\code{@use}),
neither it nor its operand \code{\%a} can be eliminated, forcing the newly introduced \code{add} to remain in the \ir.
An existing fold then fires on the
\code{add}-indexed \code{getelementptr},
reverting \code{(i8*)y+(add a,8148)}
back to the original \code{((i8*)y+a)+8148}.
These conflicting rewrites cycle endlessly,
creating an infinite optimization loop.

\begin{figure}[t]
    \TSCodeInput{figures/rq3-one-use-nc-input.ts.tex}
    \caption{Non-convergence caused by missing one-use guard on \llvmPRURL{170439}.}
    \label{fig:rq3:cs:one-use-nc}
\end{figure}

\paragraph{Poison/FMF Flag Misuse} \label{par:rq3:flag-misuse}

Flag misuse arises when the patch incorrectly adds or drops poison-related
or fast-math flags (FMF), or omits a flag check in the match condition.
Dropping a flag blocks downstream folds that depend on it, causing a missed optimization.
Adding a flag without justification is unsound: for \irs that violate the flag's precondition,
the target produces poison due to the new flag where the source does not.
Omitting a flag guard is equally unsound: the fold fires without verifying the flag's presence,
yet the rewrite's correctness relies on the guarantee it provides.
For example, the \code{samesign} flag on \code{icmp} assumes
that both operands share the same sign. 
An optimization might leverage this flag
to rewrite \code{icmp samesign ult x, y} into \code{icmp slt x, y}.
Omitting the flag check allows the fold to unsoundly apply to an ordinary \code{icmp ult},
which evaluates differently for sign-mixed inputs
(\eg, \code{-1 ult 1} is 0, while \code{-1 slt 1} is 1).

\paragraph{Unsound Transform}

The patch may implement a semantically incorrect transformation, making this the
most severe failure mode as it can lead to miscompilation.
Unlike the poison/FMF flag theme, the root causes here are logical errors
in the rewrite itself, \eg, missing a guard on constant ranges
or incorrectly swapping the two arms of a \code{select}.

\paragraph{Non-convergent Rewrite}

A non-convergent rewrite prevents \instcombine from reaching a fixpoint,
triggering an assertion failure or an infinite loop
(note that non-convergence caused by missing one-use guards,
such as in \cref{fig:rq3:cs:one-use-nc},
is categorized under \emph{incorrect one-use guard}).
These failures are severe because they can hang the compiler indefinitely.
In the current \llvm (version 22 as of writing), \instcombine
bounds its outer iterations but leaves the inner worklist loop unbounded.
Consequently, conflicting rewrites that continually spawn new instructions
cycle endlessly within the worklist and hang the compiler (\emph{timeout}),
whereas cycles spanning consecutive \instcombine iterations
are safely caught by the outer limit (\emph{unreachable fixpoint}).
This architectural distinction accounts for the two subcategories in \cref{tab:rq3-agent-failures}.

\myfinding[finding:rq3.1]{
    While \emph{under-generalization} and \emph{incorrect one-use guard} 
    are the most prevalent themes, \emph{unsound transform} 
    and \emph{non-convergent rewrite} are the most severe.
    These expose two core limitations: failing to 
    correctly generalize beyond concrete examples, 
    and mishandling \llvm-specific features 
    (\eg, one-use guards, flags, fold interactions) not surfaced by seed tests.
}

\subsection{Threats to Validity} \label{subsec:threats}

\myparagraph{Profitability Across Backend Architectures}
We estimate profitability with \llvmmca on the \xEightSix backend;
profitability on other architectures may differ.
This is a construct-validity threat, as our criterion proxies backend
performance rather than measuring all targets.
Since our focus is on target-independent optimizations on \llvm \ir,
the impact of backend differences is limited.

\myparagraph{Limitations of \alivetwo}
\alivetwo does not support all \llvm \ir features and may return
inconclusive results or time out on complex cases,
so our validity criterion depends on the rewrites with conclusive validation;
we mitigate this by excluding \prs where more than 10\% of test cases
are inconclusive.

\myparagraph{Randomness in Agent Outputs}
Agent outputs are non-deterministic,
so our evaluation may be sensitive to the outputs of individual runs.
We therefore run each task three times and report Pass@1 and Pass@3~\cite{chen2021evaluating}.

\myparagraph{Data Leakage}
All \prs in \proj are merged after \prCollectionStartDate,
postdating every evaluated model's knowledge cutoff.
Auditing all agent transcripts, we find that
no run accessed the solution \pr or its linked issues.

\section{Implications and Future Directions}
\label{sec:implications}
\myparagraph{Integrate Robust Validation into Agent Loops}
Mutation-based test generation increases coverage
beyond the original \filecheck-based regression tests
and reveals more failures (\cref{par:rq1:case-studies:coverage}),
and behavioral validation provides a more accurate evaluation (\cref{par:rq1:case-studies:fragile}).
Future work could integrate both into the agent's patch generation loop
to enable self-correction.

\myparagraph{Close Generalization and Compiler-Knowledge Gaps}
In RQ3 (\cref{subsec:rq3}), under-generalization is the most common failure;
moreover, profitability rates in RQ1 (\cref{subsec:rq1}) remain well below the human baseline,
and many failures stem from mishandling \llvm-specific features
that existing seed tests do not expose.
Future work could tailor mutation-based test generation to target these
features more aggressively, and explore prompting or training strategies
that ask agents to enumerate edge cases explicitly or incorporate
\llvm's pattern-matching grammar and feature conventions.

\myparagraph{Use Agents as a Complement to Human Experts}
RQ2 (\cref{subsec:rq2}) shows that human patches are not always behaviorally valid,
and agents can discover profitable generalizations that human patches miss,
suggesting agents could help review human patches,
flagging regressions or missed cases.

\myparagraph{Improve \ir Profitability Measurement}
Canonicalization issues (\cref{par:rq3:canonicalization})
show nearly identical \ir snippets can lower to
machine code with different profitability,
highlighting the need for robust \ir metrics.

\myparagraph{Encourage Refactoring over Rule Accumulation}
Agents produce larger patches than humans
($+$27.1--43.9 net added lines \vs\ $+$18.1),
suggesting a preference for appending rules over refactoring existing folds.
To encourage refactoring, future work could provide tool support
to help agents locate relevant folds and insertion points.

\myparagraph{Prioritize Issues by Measuring Real-World Impact}
As observed in RQ2 (\cref{subsec:rq2}), human-written patches can introduce
profitability regressions.
However, these regressions often persist without being reported by users,
suggesting that their real-world impact may be limited.
Since compilers face numerous missed optimizations, regressions,
and bugs---many of which rarely occur in practice---prioritization
is critical.
Future work should measure \ir pattern prevalence to guide efforts
toward the most impactful issues.

\section{Related Work}
\label{sec:related-work}

\myparagraph{Agentic Coding Benchmarks}
Coding agents have been evaluated on code generation in function level
(\eg HumanEval~\cite{chen2021codex}), resolving real-world GitHub issues
(\eg SWE-bench~\cite{jimenez2023swe}, SWE-Lancer~\cite{miserendino2025swe}),
and complex open-ended development tasks
(\eg TerminalBench~\cite{merrill2026terminalbench}, FeatureBench~\cite{zhou2026featurebench}).
Zheng \etal~\cite{zheng2026agentic} evaluate agents on automated
repair of real-world \llvm issues but do not assess the correctness or
profitability of the optimizations;
\proj instead targets missed \instcombine peephole optimizations
and evaluates patches behaviorally.

\myparagraph{\llm-Based Approaches for Code Optimization}
Several studies explore \llms for code and compiler
optimization~\cite{garg2023rapgen,gao2024search,cummins2023llm,grubisic2024compiler,fang2024towards,italiano2024finding},
some at the source level only~\cite{garg2023rapgen,gao2024search}.
Cummins \etal~\cite{cummins2023llm} and Grubisic \etal~\cite{grubisic2024compiler} train \llms to predict compiler flags for \llvm code-size reduction.
Fang \etal~\cite{fang2024towards} test whether \llms can learn a single \aarchSixFour peephole rule.
Italiano and Cummins~\cite{italiano2024finding} use \llms to discover missed code-size optimizations but without implementing or verifying the rewrites.
Yang \etal~\cite{yang2026ir} construct an optimization-sensitive \ir dataset for \llm-based \ir optimizers.
Qiu \etal~\cite{qiu2026beyond} propose intent-driven \ir optimization where \llms apply transformations guided by high-level intent.
Most closely related, LPO~\cite{xu2026lpo} uses \llms to \emph{discover} missed
\instcombine optimizations through an \alivetwo-guided feedback loop.
\proj instead evaluates coding agents on \emph{implementing} missed optimizations
from real issue reports, and assesses patches along correctness, profitability,
and generalization dimensions.

\section{Conclusion}
\label{sec:conclusion}
We presented \proj, a benchmark of \numPRs missed \instcombine peephole
optimizations from real \llvm \prs.
Our evaluation reveals a trade-off between correctness and profitability:
no agent simultaneously achieves human-level performance on both.
Failure modes center on under-generalization
and mishandling of \llvm-specific conventions
that existing test suites largely miss.

\bibliographystyle{IEEEtran}
\bibliography{main}



\end{document}